\documentclass[useAMS,usenatbib]{mn2e}
\usepackage{graphicx}
\usepackage{rotating}
\usepackage{longtable}
\usepackage{lscape}

\def\Msun{M$_\odot$}
\def\Lsun{L$_\odot$}
\def\Mbh{${\cal M}_{\rm BH}$}

\def\Mhost{${\cal M}_{\rm host}$}

\def\Civ{C\,{\sc iv}}
\def\Mgii{Mg\,{\sc ii}}
\def\FeII{Fe\,{\sc ii}}

\def\Oiii{[O\,{\sc iii}]}
\def\Hb{H$\beta$}
\def\Ha{H$\alpha$}
\def\lsim{\mathrel{\rlap{\lower 3pt \hbox{$\sim$}} \raise 2.0pt \hbox{$<$}}}
\def\gsim{\mathrel{\rlap{\lower 3pt \hbox{$\sim$}} \raise 2.0pt \hbox{$>$}}}

\title[The broad line region in blazars]{Geometry and inclination of the broad line region in blazars}
\author[R. Decarli et al.]{
R. Decarli$^{1}$\thanks{E-mail: decarli@mpia.de},
M. Dotti$^{2}$ and
A. Treves$^{3}$ 
\\
$^{1}$Max-Planck-Institut f\"ur Astronomie, K\"onigstuhl 17, 69117 Heidelberg, Germany\\
$^{2}$Max-Planck-Institut f\"{u}r Astrophysik, Karl-Schwarzschild-Str. 1, D-85748 Garching, Germany\\
$^{3}$Dipartimento di Fisica e Matematica, Universit\`a dell'Insubria, via Valleggio 11, I-22100 Como, Italy
}

\begin{document}

\date{}

\pagerange{\pageref{firstpage}--\pageref{lastpage}} \pubyear{}

\maketitle

\label{firstpage}

\begin{abstract}
We study the properties of the broad line region in blazars by
comparing the virial estimate of black hole masses with that derived
from the mass of the host galaxies. The former is sensitive to the
width of broad lines, i.e., to the projection of the velocity of
line-emitting clouds along the line of sight; the latter is not.
This comparison allows us to constrain the deprojection factor $f$,
thus revealing general properties of the geometry of the broad line
region. We show that blazars tend to have 1) higher $f$ values than the
quasars of our reference sample: $\langle f_{\rm BLLacs}\rangle=6.9\pm2.3$,
$\langle f_{\rm blazars}\rangle=5.6\pm1.3$ and
$\langle f_{\rm quasars}\rangle=2.0\pm0.3$; 2) relatively narrow broad emission
lines; 3) modest equivalent widths, as expected because of the
occurrence of jet emission at very low inclination angles. In a
disc-like sketch of the broad line region, these results indicate a
pole-on view of a flat geometry in blazars. This consistently extends
the orientation--dependent unified model of active nuclei to the
geometry of the broad line region.
\end{abstract}


\section{Introduction}

The estimate of black hole masses (\Mbh) in distant galaxies is
feasible only in active galactic nuclei (AGN), where we can infer
the dynamics of gas close to the black holes (BHs) through its
bright continuum and line emission. A common practice for type-1 AGN
(where broad emission lines are observed) consists in adopting the
virial paradigm:
\begin{equation}\label{eq_virial}
{\cal M}_{\rm BH}=\frac{R_{\rm BLR} v_{\rm BLR}^2}{G}
\end{equation}
where $R_{\rm BLR}$ and $v_{\rm BLR}$ are respectively the orbital
radii and the velocity of the gas clouds in the broad line region
(BLR), and $G$ is the gravitational constant. The cloud velocity is
usually inferred from the width of broad lines, assuming a
deprojection factor $f$:
\begin{equation}\label{eq_fdef}
v_{\rm BLR}=f \cdot {\rm FWHM}
\end{equation}
while $R_{\rm BLR}$ is derived from the continuum luminosity of the
accretion disc \citep{kaspi00,kaspi05,kaspi07,bentz09}. This allows an
estimate of \Mbh{} from single-epoch spectra: \citet{shen10} applied
this technique to the huge spectroscopic sample of quasars in the
Sloan Digital Sky Survey (SDSS), thus providing \Mbh{} for $\sim
100,000$ sources.

The virial technique suffers of a number of limitations. In
particular, the deprojection value $f$ is not known, and is strongly
sensitive to the (unknown) geometry and orientation of the BLR.
In the commonly assumed isotropic picture of the BLR,
$\langle f \rangle = \sqrt{3/4}$ \citep{netzer90}.
Assuming instead a disc--like geometry \citep[e.g.,][]{mclure01},
\begin{equation}\label{eq_disc}
f=0.5~\left[ \left( \frac{H}{R} \right)^2+\sin^2 \vartheta\right]^{-1/2},
\end{equation}
where $H/R$ is the aspect ratio of the disc and $\vartheta$ is the angle
between the normal to the disc and the line of sight. In the limit 
of geometrically thin discs, equation~\ref{eq_disc} reduces to 
$f=0.5 / (\sin \vartheta)$, ranging between 0.5 (edge-on discs)
and $\infty$ (face-on discs). Taking into account the thickness
of the disc, $f \rightarrow 0.5~(H/R)^{-1}$ when $\vartheta \rightarrow 0$.
For disc--like BLRs, $\langle f \rangle\approx 1$ almost independently
of $H/R$ \citep{nls1}.

From an observational point of view, average values for $f$ have been 
estimated by comparing the mean offset of \Mbh{} estimates based on 
different techniques. \citet{onken04} compared the \Mbh{} values expected 
from the \Mbh{}--$\sigma_*$ relation for a set of local type-1 AGN with those 
obtained from the FWHM of \Hb{} through equation \ref{eq_virial}, finding 
$\langle f ($H$\beta)\rangle = 2.3$\footnote{The average value reported here
has been recomputed to account for the different definition of $f$ used
in \citet{onken04}.}. 
\citet{collin06} repeated the same exercise for a larger set of objects, 
and found a dependence of $\langle f ($H$\beta)\rangle$ on the broad line 
width ($\langle f \rangle\approx 1.5$ for FWHM$<$2000 km/s, $\langle f 
\rangle\approx 0.7$ for FWHM$>$6000 km/s), which could be attributed to the 
Eddington factor or to the BLR orientation.
\citet{decarli08} expanded on the previous results by \citet{labita06},
comparing virial estimates of \Mbh{} based on \Hb{} and \Civ{} lines
with those inferred from the host galaxy luminosities assuming the
scaling in \citet{bettoni03}. They found $\langle f ($H$\beta)\rangle
\approx 1.6$, $\langle f($\Civ{}$)\rangle\approx2.4$ and a strong
dependence of $f$ on the widths of broad lines, i.e., $f>2$ values are found
only in quasars with FWHM$<$5000 km/s \citep[see also][]{lamura07,lamura09}.
The different values of $\langle f \rangle$ for different broad emission lines
can be explained by a different geometry of the corresponding emitting
regions \citep[see also][]{fine08,fine10}.

\citet{nls1} noticed that the claimed ``black hole mass deficit'' observed in
Narrow-Line Seyfert 1 galaxies \citep[NLS1s; see][]{grupe04} 
disappears once a disc--like BLR is assumed. NLS1s show relatively narrow permitted
lines (500 km/s $<$ FWHM $<$ 2000 km/s), small \Mbh{} with respect
to the predictions from the \Mbh{}--host galaxy relations, and high
Eddington ratios. Assuming that the BLR is flat and seen almost
pole-on, from the statistics of NLS1s with respect to `normal'
Seyfert 1 galaxies we can infer the average $f$ value for NLS1s.
Interestingly, once corrected for this higher deprojection values,
line widths yield to distributions of \Mbh{} and Eddington ratios 
almost indistinguishable from normal
Seyfert 1 galaxies. The strong radio loudness with blazar-like
properties observed in a large fraction of NLS1 \citep{yuan08} and
the recent detection of GeV emission in a handful of NLS1s
\citep{abdo09a,abdo09b} also reveal the presence of beamed emission
almost aligned with the line of sight, supporting a pole-on picture
of the central engine of these AGN. 

Other indications supporting a flat geometry for the BLR come from
1) the anti-correlation between the width of broad emission lines and
the ``compactness'' parameter defined from radio observations as the
luminosity ratio between the core and the lobe emission in radio
galaxies and radio-loud quasars
\citep{wills86,brotherton96,vestergaard00}; 2) the dependence of the
FWHM of broad lines on the Equivalent Width of the \Oiii{} line
\citep{risaliti10}.

In order to test the flat geometry picture of the BLR, we here
extend our previous studies to blazars. According to the unified
model of AGN \citep{antonucci85,urry95}, blazar jets (and specifically
BL Lac objects) are expected to show small inclination angles with 
respect to the line of sight. If the BLR has a flat geometry with a 
rotational axis aligned to the jet, blazars should show relatively
narrow FWHM of broad lines and high values of $f$. We exploit a newly 
defined recipe to estimate the radius of the BLR from the luminosity of
broad lines \citep[see also][]{vestergaard06} in order to avoid the 
contamination from the jet light in the optical continuum emission. 
Our virial estimates of \Mbh{} are then compared to the
estimates we have from the \Mbh--\Mhost{} relation in order to
derive $f$.

This paper is organized as follows: In section \ref{sec_sample} we
present our sample. Our analysis is described in section
\ref{sec_analysis}, while in section \ref{sec_results} we present
our observational results and compare them with expectations from
simple BLR models. Conclusions are drawn in section \ref{sec_conclusions}.
Throughout the paper we will assume a concordance cosmology with
$H_0=70$ km/s/Mpc, $\Omega_{\rm m}=0.3$ and $\Omega_{\rm
\Lambda}=0.7$. Results from other papers are accordingly scaled to
this cosmology.

\section[]{The sample}\label{sec_sample}

In order to perform our analysis, we require an estimate of the host
galaxy luminosity and a characterization of the properties (flux,
width) of at least one broad line.
The blazar sample is mostly derived from the compilations by
\citet{scarpa00}, \citet{kotilainen98} and \citet{nilsson03}. Few
other sources are taken from \citet{wurtz96}, \citet{nilsson09} and 
\citet{decarli10b}. We select all the blazars with a reliable estimate 
of the host galaxy luminosity. We adopt published
values of $M_{\rm R}$(host) or convert available bands into $R$
computing the $k$-correction through the Elliptical galaxy template
by \citet{mannucci01}. Out of this selection, we pick up only those
objects with a broad line detected in an optical spectrum. Main data
sources for the spectroscopy are the {\sf zbllac}
database\footnote{See \textsf{http://www.oapd.inaf.it/zbllac/}.} and
the SDSS archive. For BL Lac and OJ287, we use the FWHM and flux
values published by \citet{capetti10} and \citet{nilsson10}
respectively. It turns out that the only two lines of interest are
\Mgii{}$_{\lambda 2798}$ and \Ha{}. The former is the only broad
line observed at $0.5<z<1.6$. We preferred the latter rather than
\Hb{} because it is $\sim3\times$ brighter. The whole blazar sample
consists of 18 objects.

In order to understand the role of orientation of the broad line
region with respect to `normal' quasars, we include a comparison
sample of quasars from the data set presented in
\citet{decarli10a,decarli10b}. We consider only objects where the
\Mgii{} or \Ha{} lines have been fully covered in the observed
wavelength range. The reference sample consists of 39 sources. Table
\ref{tab_sample} summarizes the objects in both samples.

\begin{table*}
  \caption{The sample. (1) Source name. (2--3) Source coordinate
  in J2000. (4) Redshift. (5) Absolute rest-frame $R$-band magnitude of
  the host galaxy. (6) Adopted Mass to Light ratio, in solar units. (7)
  \Mbh{} derived from the host galaxy luminosity. (8) Reference for the
  host galaxy luminosity: a- \citet{kotilainen98}; b- \citet{scarpa00}; 
  c- \citet{nilsson03}; d- \citet{wurtz96}; e- \citet{decarli10b};
  f- \citet{nilsson09}.
}\label{tab_sample}
	\centering
\small
\begin{tabular}{cccccccc}
	\hline
   Name	&  R.A.      &  Dec.	 & $z$ & $M_{\rm R}$ (host) & ${\cal M}/L$ & log \Mbh(ima) & Ref. \\
		&	     &  	 &     & [mag] & [\Msun/\Lsun] & [\Msun] &  \\
    (1)	&  (2)       &  (3)	 & (4) & (5)   & (6) & (7) & (8) \\
 \hline
 \multicolumn{8}{l}{{\it Blazars}}\\												 
   PKS0403-132 & 04:05:34.0 & -13:08:14 &0.571& -21.3 &0.60&  8.3  & a \\	
   PKS0405-123 & 04:07:48.5 & -12:11:36 &0.574& -23.1 &0.60&  9.0  & a \\	
   PKS0420-014 & 04:23:15.8 & -01:20:33 &0.915& -24.5 &0.51&  9.5  & a \\	
   PKS0426-380 & 04:28:40.4 & -37:56:20 &1.105& -23.2 &0.44&  8.9  & b \\	
    PKS0521-36 & 05:22:58.0 & -36:27:31 &0.055& -22.4 &0.78&  8.9  & b \\	
 	 OJ049 & 08:31:48.9 & +04:29:39 &0.174& -23.4 &0.73&  9.3  & c \\	
 	 OJ287 & 08:54:48.9 & +20:06:31 &0.306& -22.0 &0.68&  8.7  & d \\	
 	 3C273 & 12:29:06.7 & +02:03:09 &0.158& -22.8 &0.74&  9.1  & e \\	
      1249+174 & 12:51:45.3 & +17:11:18 &0.644& -21.0 &0.58&  8.2  & b \\	
 	 3C279 & 12:56:11.1 & -05:47:22 &0.538& -24.0 &0.61&  9.4  & f \\
   PKS1519-273 & 15:22:37.7 & -27:30:11 &1.297& -24.4 &0.38&  9.3  & b \\
 	4C1460 & 15:40:49.5 & +14:47:46 &0.605& -24.3 &0.59&  9.5  & c \\
 	 3C345 & 16:42:58.8 & +39:48:37 &0.594& -25.5 &0.60& 10.0  & e \\
 	 3C371 & 18:06:50.7 & +69:49:28 &0.051& -22.6 &0.78&  9.0  & c \\
   PKS2131-021 & 21:34:10.3 & -01:53:17 &1.284& -22.8 &0.39&  8.7  & b \\
 	BL Lac & 22:02:43.3 & +42:16:40 &0.069& -21.9 &0.78&  8.7  & b \\
    PKS2201+04 & 22:04:17.6 & +04:40:02 &0.027& -21.6 &0.79&  8.6  & b \\
   PKS2345-167 & 23:48:02.6 & -16:31:13 &0.576& -24.4 &0.60&  9.6  & a \\
 \hline
 \multicolumn{8}{l}{{\it Reference quasar sample}}\\
   PKS0000-177 & 00:03:22.0 & -17:27:12 &1.465& -24.5 &0.33&  9.3  & e \\	 
       SGP5:46 & 00:52:22.8 & -27:30:03 &0.955& -22.4 &0.50&  8.7  & e \\	 
      0054+144 & 00:57:09.9 & +14:46:10 &0.171& -22.9 &0.74&  9.1  & e \\	 
   PKS0100-270 & 01:02:56.3 & -26:46:36 &1.597& -23.4 &0.30&  8.9  & e \\	 
      0119-370 & 01:21:24.1 & -36:50:02 &1.320& -23.7 &0.37&  9.1  & e \\	 
      0133+207 & 01:36:24.4 & +20:57:27 &0.425& -22.8 &0.65&  9.0  & e \\	 
     0152-4055 & 01:54:20.1 & -40:40:30 &1.650& -23.4 &0.29&  8.8  & e \\	 
   PKS0155-495 & 01:57:38.0 & -49:15:19 &1.298& -24.4 &0.38&  9.3  & e \\	 
    PKS0159-11 & 02:01:57.1 & -11:32:34 &0.669& -22.3 &0.57&  8.7  & e \\	 
 	PB6708 & 02:09:48.8 & -00:43:36 &0.868& -22.8 &0.52&  8.8  & e \\	 
 KUV03086-0447 & 03:11:04.7 & -04:35:41 &0.755& -23.7 &0.55&  9.2  & e \\	 
 	US3828 & 03:18:25.6 & +15:59:56 &0.515& -22.6 &0.62&  8.9  & e \\	 
   PKS0348-120 & 03:51:11.0 & -11:53:23 &1.520& -24.8 &0.32&  9.4  & e \\	 
    PKS0349-14 & 03:51:28.6 & -14:29:10 &0.614& -25.2 &0.59&  9.9  & e \\	 
   PKS0402-362 & 04:03:53.8 & -36:05:02 &1.417& -24.8 &0.34&  9.5  & e \\	 
    PKS0414-06 & 04:17:16.7 & -05:53:45 &0.773& -25.0 &0.54&  9.7  & e \\	 
    PKS0440-00 & 04:42:38.6 & -00:17:43 &0.607& -23.3 &0.59&  9.1  & e \\	 
     0624+6907 & 06:30:02.5 & +69:05:04 &0.370& -24.6 &0.66&  9.7  & e \\	 
    PKS0710+11 & 07:13:02.3 & +11:46:15 &0.768& -25.6 &0.54& 10.0  & e \\	 
  MS07546+3928 & 07:58:00.0 & +39:20:29 &0.096& -22.0 &0.77&  8.8  & e \\	 
 MS0824.2+0327 & 08:26:52.9 & +03:17:13 &1.431& -23.8 &0.33&  9.1  & e \\	 
  MS08287+6614 & 08:33:17.9 & +66:03:46 &0.610& -22.9 &0.59&  9.0  & e \\	 
    PKS0838+13 & 08:40:47.6 & +13:12:23 &0.684& -23.1 &0.57&  9.0  & e \\	 
 	US1867 & 08:53:34.2 & +43:49:02 &0.513& -25.8 &0.62& 10.1  & e \\	 
 	TON392 & 09:12:17.8 & +24:50:38 &0.654& -23.4 &0.58&  9.1  & e \\	 
  MS09441+1333 & 09:46:52.0 & +13:20:26 &0.131& -23.3 &0.75&  9.3  & e \\	 
      1004+130 & 10:07:26.1 & +12:48:56 &0.240& -23.6 &0.71&  9.3  & e \\	 
  Z101733-0203 & 10:17:33.5 & -02:03:07 &1.343& -21.8 &0.36&  8.3  & e \\	 
    PKS1015-31 & 10:18:09.3 & -31:44:14 &1.346& -24.5 &0.36&  9.4  & e \\	 
    PKS1018-42 & 10:20:03.9 & -42:51:30 &1.280& -25.1 &0.39&  9.6  & e \\	 
      1150+497 & 11:53:24.4 & +49:31:09 &0.334& -23.4 &0.68&  9.2  & e \\	 
      1402+436 & 14:04:38.8 & +43:27:07 &0.320& -23.3 &0.68&  9.2  & e \\	 
      1444+407 & 14:46:45.9 & +40:35:06 &0.267& -22.3 &0.70&  8.8  & e \\	 
   PKSJ1511-10 & 15:13:44.9 & -10:12:00 &1.513& -23.5 &0.32&  8.9  & e \\	 
   HS1623+7313 & 16:22:16.8 & +73:06:15 &0.621& -22.2 &0.59&  8.7  & e \\	 
    MC2112+172 & 21:14:56.7 & +17:29:23 &0.878& -24.1 &0.52&  9.4  & e \\	 
    Q2225-403B & 22:28:50.4 & -40:08:27 &0.932& -23.2 &0.50&  9.0  & e \\	 
      2247+140 & 22:50:25.3 & +14:19:52 &0.235& -23.0 &0.71&  9.1  & e \\	 
    Q2348-4012 & 23:51:02.1 & -39:56:18 &1.500& -22.1 &0.32&  8.4  & e \\	 
      \hline
      \end{tabular}
\end{table*}

\section[]{Data analysis}\label{sec_analysis}					
										 
\subsection{Measured quantities}						 
\begin{figure}									 
\begin{center}									 
\includegraphics[width=0.49\textwidth]{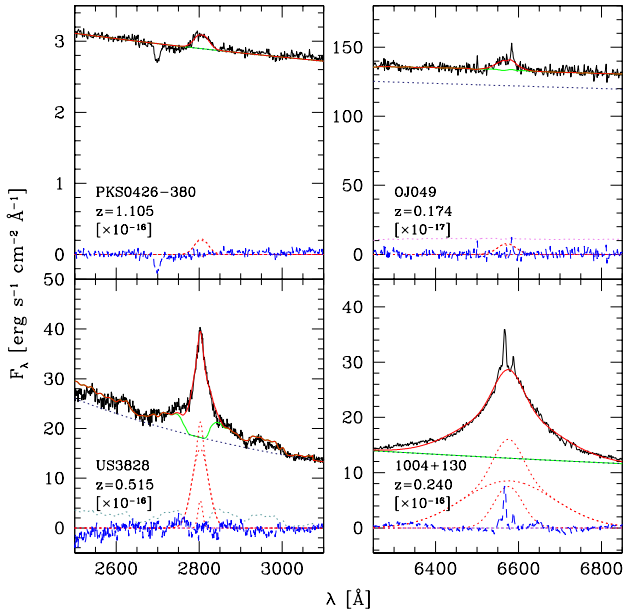}\\				
\caption{Examples of the continuum and line fits. The spectra of two blazars	
(PKS0426-380 and OJ049) and two quasars (US3828 and 1004+130) are shown. The	
solid black lines show the observed spectra, the solid red lines mark the	    
model, which results from the superposition of a power-law continuum		    
(dotted black lines), the iron and host galaxy templates (dotted light blue 	    
lines) and a line model (dotted red lines). The residuals after model 		    
subtraction are shown as blue dashed lines (a colour version of this plot is	
available in the electronic edition of the journal).}\label{fig_spec}		    
\end{center}
\end{figure}

\begin{table*}
  \caption{Results from spectroscopy. (1) Source name. 
  (2) Line used to estimate \Mbh{}. (3) FWHM. (4) Line luminosity.
  (5) Power-law monochromatic luminosity at 3000 \AA{} (for
  \Mgii{}) or 5100 \AA{} (for \Ha{}). (6) Line Equivalent Width. (7)
  \Mbh{} estimate based on the line luminosity and width. (8) Geometrical
  factor $f$. (9) Reference for the spectra:
  a- \citet{decarli10a};
  b- \citet{sbarufatti05}; 
  c- \citet{sbarufatti06aa}; 
  d- SDSS archive; 
  e- \citet{nilsson10}; 
  f- \citet{labita06}; 
  g- \citet{pian05}; 
  h- HST STIS archive; 
  i- \citet{sbarufatti06aj}; 
  j- \citet{capetti10}.}\label{tab_spec}
	\centering
\small
\begin{tabular}{ccccccccc}
	\hline
   Name	& Line & FWHM & log $L_{\rm line}$ & log $\lambda L_\lambda$ & E.W. & log \Mbh{}(spec) & $f$ & Ref. \\
		& & [km/s] & [erg/s] & [erg/s]  &  [\AA{}] & [\Msun] &  & \\
    (1)	&  (2)       &  (3)	 & (4) & (5)   & (6) & (7) & (8) & (9)  \\
 \hline
 \multicolumn{9}{l}{{\it Blazars}}\\												 
   PKS0403-132 &  \Mgii &  3451 & 43.86 & 45.77 &   47.5 &  8.6 &  0.6 & a \\ 
   PKS0405-123 &  \Mgii &  3933 & 44.49 & 46.40 &   14.6 &  9.0 &  0.9 & a \\ 
   PKS0420-014 &  \Mgii &  3607 & 43.92 & 45.83 &   18.8 &  8.7 &  2.2 & a \\ 
   PKS0426-380 &  \Mgii &  4419 & 42.76 & 44.67 &    3.0 &  8.3 &  1.7 & b \\ 
    PKS0521-36 &  \Ha	&  4657 & 41.71 & 42.94 &   51.9 &  7.2 &  7.0 & c \\ 
 	 OJ049 &  \Ha	&  2553 & 41.55 & 42.78 &    3.5 &  6.6 & 21.3 & d \\ 
 	 OJ287 &  \Ha	&  3710 & 41.83 & 43.06 &    0.5 &  7.1 &  5.6 & e \\ 
 	 3C273 &  \Mgii &  4074 & 43.88 & 45.79 &   22.4 &  8.8 &  1.4 & f \\ 
      1249+174 &  \Mgii &  3040 & 43.01 & 44.91 &   27.5 &  8.1 &  1.0 & d \\ 
 	 3C279 &  \Mgii &  6534 & 43.29 & 45.20 &    8.6 &  8.9 &  1.6 & g \\ 
   PKS1519-273 &  \Mgii &  1821 & 42.37 & 44.28 &    0.4 &  7.4 &  7.9 & b \\
 	4C1460 &  \Mgii &  2697 & 41.86 & 43.76 &    1.1 &  7.5 &  9.6 & d \\
 	 3C345 &  \Mgii &  4628 & 42.77 & 44.68 &   13.5 &  8.4 &  6.0 & a \\
 	 3C371 &  \Ha	&  2333 & 42.83 & 44.07 &   14.9 &  7.4 &  6.5 & h \\
   PKS2131-021 &  \Mgii &  3668 & 42.61 & 44.52 &    2.5 &  8.1 &  1.7 & i \\
 	BL Lac &  \Ha	&  4600 & 41.59 & 42.82 &    9.9 &  7.2 &  6.1 & j \\
    PKS2201+04 &  \Ha	&  3556 & 40.61 & 41.84 &   18.8 &  6.3 & 15.0 & c \\
   PKS2345-167 &  \Mgii &  3068 & 43.40 & 45.31 &   20.3 &  8.3 &  3.9 & a \\
 \hline 								
 \multicolumn{9}{l}{{\it Reference quasar sample}}\\			
   PKS0000-177 &  \Mgii &  6475 & 43.74 & 45.65 &   19.9 &  9.1 &  1.0 & a \\
       SGP5:46 &  \Mgii &  3504 & 42.94 & 44.85 &   23.4 &  8.2 &  1.4 & a \\
      0054+144 &  \Ha	&  8635 & 42.53 & 43.76 &  399.0 &  8.3 &  2.4 & a \\
   PKS0100-270 &  \Mgii &  3384 & 44.22 & 46.12 &   15.5 &  8.8 &  0.9 & a \\
      0119-370 &  \Mgii &  6229 & 43.58 & 45.49 &   60.3 &  9.0 &  0.9 & a \\
      0133+207 &  \Mgii & 12408 & 43.72 & 45.62 &   77.4 &  9.7 &  0.4 & a \\
     0152-4055 &  \Mgii &  4924 & 43.91 & 45.82 &   31.1 &  9.0 &  0.7 & a \\
   PKS0155-495 &  \Mgii &  5395 & 43.81 & 45.72 &   50.9 &  9.0 &  1.2 & a \\
    PKS0159-11 &  \Mgii &  4140 & 44.17 & 46.08 &   26.4 &  8.9 &  0.7 & a \\
 	PB6708 &  \Mgii &  3156 & 43.58 & 45.48 &   53.8 &  8.4 &  1.4 & a \\
 KUV03086-0447 &  \Mgii &  2441 & 43.94 & 45.84 &   19.0 &  8.4 &  2.4 & a \\
 	US3828 &  \Mgii &  3779 & 43.95 & 45.85 &   47.2 &  8.7 &  1.1 & a \\
   PKS0348-120 &  \Mgii &  3699 & 43.92 & 45.83 &   27.3 &  8.7 &  1.8 & a \\
    PKS0349-14 &  \Mgii &  5841 & 44.21 & 46.12 &   21.6 &  9.2 &  1.9 & a \\
   PKS0402-362 &  \Mgii &  4619 & 44.19 & 46.10 &   17.8 &  9.0 &  1.3 & a \\
    PKS0414-06 &  \Mgii &  5732 & 44.38 & 46.29 &   45.7 &  9.3 &  1.5 & a \\
    PKS0440-00 &  \Mgii &  5585 & 43.29 & 45.20 &   23.7 &  8.8 &  1.3 & a \\
     0624+6907 &  \Mgii &  3989 & 44.50 & 46.40 &   24.5 &  9.0 &  2.0 & a \\
    PKS0710+11 &  \Mgii & 11759 & 44.30 & 46.21 &   37.7 &  9.9 &  1.0 & a \\
  MS07546+3928 &  \Ha	&  2126 & 43.52 & 44.75 &  362.7 &  7.8 &  3.1 & a \\
 MS0824.2+0327 &  \Mgii &  2812 & 43.76 & 45.66 &   29.6 &  8.4 &  1.7 & a \\
  MS08287+6614 &  \Mgii &  6403 & 43.54 & 45.44 &   32.8 &  9.0 &  0.9 & a \\
    PKS0838+13 &  \Mgii &  4383 & 43.42 & 45.32 &  132.1 &  8.6 &  1.4 & a \\
 	US1867 &  \Mgii &  2693 & 43.64 & 45.55 &   27.2 &  8.3 &  7.7 & a \\
 	TON392 &  \Mgii &  4899 & 44.34 & 46.25 &   31.5 &  9.2 &  0.9 & a \\
  MS09441+1333 &  \Ha	&  2564 & 43.12 & 44.35 &  344.6 &  7.7 &  6.2 & a \\
      1004+130 &  \Ha	&  6356 & 43.72 & 44.95 &  238.9 &  8.8 &  1.7 & a \\
  Z101733-0203 &  \Mgii &  2914 & 42.61 & 44.52 &   28.5 &  7.9 &  1.3 & a \\
    PKS1015-31 &  \Mgii &  3835 & 43.14 & 45.04 & 1001.3 &  8.4 &  2.5 & a \\
    PKS1018-42 &  \Mgii &  4105 & 44.18 & 46.09 &   34.8 &  8.9 &  1.8 & a \\
      1150+497 &  \Ha	&  3523 & 43.47 & 44.70 &  439.4 &  8.2 &  3.2 & a \\
      1402+436 &  \Ha	&  2556 & 43.01 & 44.24 &  249.5 &  7.6 &  6.1 & a \\
      1444+407 &  \Ha	&  2694 & 42.69 & 43.92 &  355.2 &  7.4 &  4.8 & a \\
   PKSJ1511-10 &  \Mgii &  3285 & 43.72 & 45.63 &   16.0 &  8.5 &  1.3 & a \\
   HS1623+7313 &  \Mgii &  3250 & 43.38 & 45.29 &   36.3 &  8.3 &  1.3 & a \\
    MC2112+172 &  \Mgii &  4221 & 43.28 & 45.19 &   38.2 &  8.5 &  2.3 & a \\
    Q2225-403B &  \Mgii &  4479 & 43.22 & 45.13 &   48.2 &  8.5 &  1.4 & a \\
      2247+140 &  \Ha	&  2832 & 42.86 & 44.10 &  120.8 &  7.6 &  5.6 & a \\
    Q2348-4012 &  \Mgii &  3316 & 44.01 & 45.92 &   39.0 &  8.7 &  0.6 & a \\
      \hline									 
      \end{tabular}								 
\end{table*}									

With the only exception of BL Lac and OJ287, for which published
values have been used \citep{capetti10,nilsson09}, the spectra of 
all the blazars and reference quasars have been fitted with the fitting 
tools developed by our group and presented in \citet{decarli10a} and
\citet{derosa10}. Namely, in order to infer the continuum and line
properties, we apply a multicomponent fit to the continuum emission
(power-law + \FeII{} model), subtract it, and fit the residual with
the sum of two Gaussian functions with the same peak wavelength
\citep[see][]{decarli08,decarli10a,derosa10}. The iron emission is
modelled through the \citet{vestergaard01} template in the UV and
our own spectrum of I Zw001 in the optical wavelengths, after
removing main emission lines and convolving to a Gaussian profile to
mimic the Doppler broadening. Free parameters in the continuum
estimate are the slope and normalization of the power-law and the
intensity of the \FeII{} feature. The iron multiplets are found to
be negligible in the wavelength range bracketing \Ha{}. Two sources
(PKS0521-36 and PKS2201+04) show significant emission from the
host galaxy in their optical spectra. We modelled this component
using the Elliptical template by \citet{mannucci01}. According to
our best fit, at 5100 \AA{}, the host galaxies of these two objects
account for $\sim30$\% of the specific flux. Two more objects
(3C371 and OJ049) also show marginal ($\lsim5$ \% of the flux density at
5100 \AA{}; see Figure \ref{fig_spec}) contribution from the host galaxies.

The line fit has 5 free parameters, namely the line peak, and the
width and amplitude of the two Gaussian curves. Figure \ref{fig_spec}
shows four examples of fitting. From the line model we compute the Full
Width at Half Maximum (FWHM), the integrated line flux ($F_{\rm
line}$), and the line Equivalent Width (EW). Following \citet{derosa10},
we compute uncertainties in the parameters by marginalizing the
distribution of values yielding $\chi^2<\chi_{\rm best}^2+1$.
Fitting errors in the luminosities (both for the line and the
continuum) are usually $\sim5$\%, and never exceed 10\%. Typical
errors on the FWHM are $\sim10$\% \citep[see also the analysis
in][]{decarli08}. We remark that these errors do not contribute
significantly to the uncertainties on \Mbh{} and $f$, which are dominated
by the scatter in the $R_{\rm BLR}$--luminosity relations ($\sim0.4$ dex).

\subsection{Derived quantities}

The $R_{\rm BLR}$--luminosity relations quoted above refer to the
continuum emission of the accretion disc. This quantity is not
directly accessible in blazars, where significant contamination from
the synchrotron emission of the jet may occur. However, quasar
spectra show a strong correlation between the luminosity of broad
lines and the continuum luminosity of the accretion disc. This is
not unexpected, as broad lines are emitted by gas directly 
photoionized by the continuum.
\citet{vestergaard06} used the broad \Hb{} line luminosity
as a proxy for the BLR radius in quasars. In the present case, this
approach benefits of avoiding contaminations by the jet emission, hence 
representing a viable tool to estimate $R_{\rm BLR}$ even in blazars. 

\begin{figure*}
\begin{center}
\includegraphics[width=0.49\textwidth]{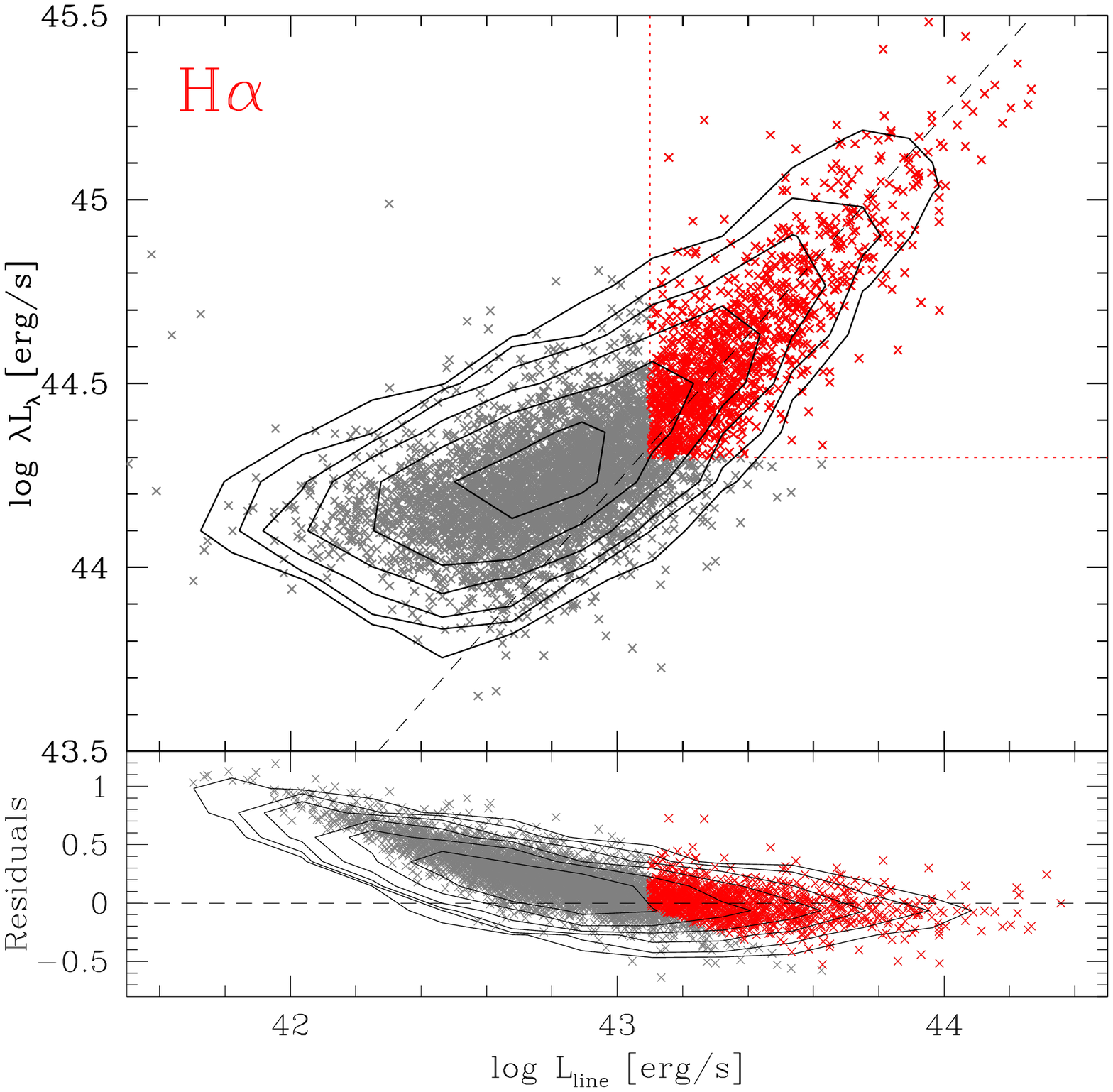}
\includegraphics[width=0.49\textwidth]{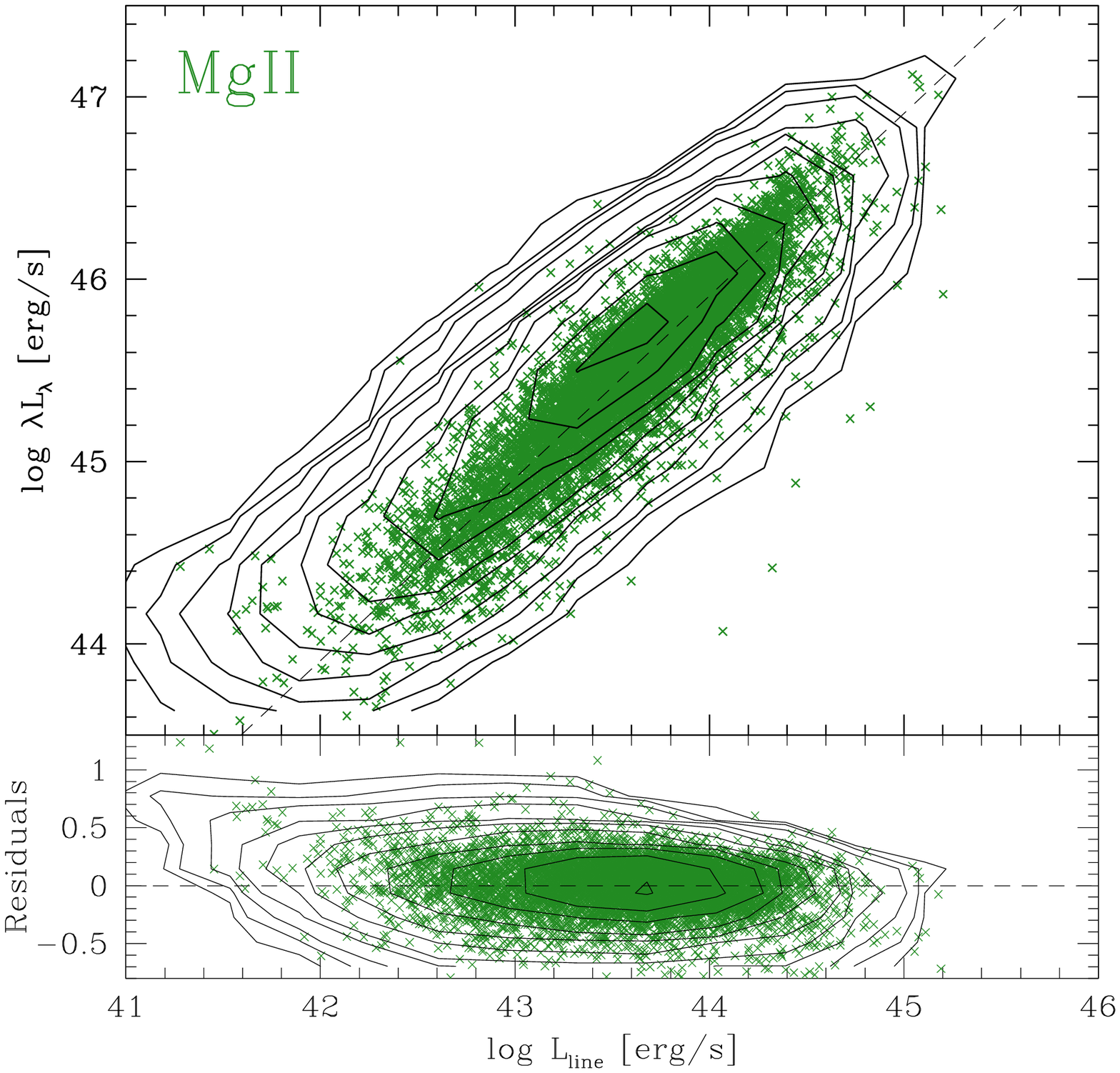}\\
\caption{The continuum to line luminosity ratios for \Ha{} 
({\it left panel}) and \Mgii{} ({\it right panel}). Data are taken from 
\citet{shen10}. For the sake of clarity, in the \Mgii{} plots, only 1 
object out of 5 is shown. Contours enclosing $>10$, 20, 40, 80, \ldots  
objects are also plotted. Dash lines show the average ratios. The plot 
region used to compute it for \Ha{} is marked with dotted lines. The faint 
end of the continuum luminosity distribution in the optical wavelengths is 
clearly deviating from the 1-to-1 
relation, because of the contribution of the host galaxy starlight. The
same is not true in the UV range, where the host galaxy contribution is
negligible throughout the luminosity range of interest, and a clear
linear correlation appears.}\label{fig_ll}
\end{center}
\end{figure*}

In Figure \ref{fig_ll} we plot the continuum versus the broad line 
luminosities in quasars from \citet{shen10} for \Ha{} and \Mgii{}. The
latter shows a clear 1-to-1 correlation spanning several order of 
magnitudes. The average ratio between the continuum and the line 
luminosities is:
\begin{equation}\label{eq_ll_m2}
\log \frac{\lambda L_{\lambda} (3000 \AA)}{L_{{\rm line}}({\rm MgII})}=1.91\pm0.26
\end{equation}
where the error is the rms, and it was computed over 85515 quasars.
Concerning \Ha{}, the scene is more complex. At low luminosities
($\lsim10^{44}$ erg/s in the continuum), significant contamination from
the host galaxy is present. \citet{bentz09} showed that the removal of this 
contribution significantly improves the quality of the $R_{\rm BLR}$--luminosity
relation in the (rest frame) optical wavelengths. We therefore focus only on
the bright end of the continuum luminosity distribution, where the role
of starlight is negligible. A similar cut is also applied to the line 
luminosity, since fits of faint lines are less reliable due to the 
contaminations from narrow \Ha{} and [N{\sc ii}] lines.
The average and rms values of the continuum to line luminosity ratios
(computed over 1162 objects) is:
\begin{equation}\label{eq_ll_ha}
\log \frac{\lambda L_{\lambda} (5100 \AA)}{L_{{\rm line}}({\rm H}\alpha)}=1.23\pm0.14.
\end{equation}
We used equations \ref{eq_ll_m2} and \ref{eq_ll_ha} to infer the disc
contribution to the continuum luminosity. Since only one of the two lines
can be observed in optical spectra (\Mgii{} or \Ha{}, according to the
target redshift), we adopted equation \ref{eq_ll_m2} for objects at $z>0.5$ 
and equation \ref{eq_ll_ha} for the remaining sources.

Once the line width, luminosity and the (corrected) continuum luminosity 
are known, we use the $R_{\rm BLR}$--luminosity relations by 
\citet{mclure02} and \citet{kaspi07} to compute \Mbh{} through equations
\ref{eq_virial}--\ref{eq_ll_ha}. These estimates are compared with
the \Mbh{} values we derive from the host galaxy. We convert host galaxy
luminosities into masses by assuming that the stellar populations are old, 
generated in a single burst of instantaneous star formation at high-$z$
\citep[following][we assume $z_{\rm burst}=5$]{decarli10b}, then
passively evolving\footnote{All the sources in our study have elliptical
host galaxies \citep{scarpa00,nilsson03,decarli10a}. Therefore, we
do not discriminate between the bulge and the whole galaxy. This is also
consistent with the assumption of an old stellar population.}. We adopt 
the mass-to-light ratio evolution predicted by \citet{bruzual03}. The 
black hole mass is then estimated through the (redshift-dependent) relation 
found in \citet{decarli10b}:
\begin{equation}\label{eq_gamma}
\log \frac{{\cal M}_{\rm BH}}{{\cal M}_{\rm bulge}} = (0.28 \pm 0.06) z - (2.91\pm0.06)
\end{equation}
We therefore can estimate $f$ from the ratio between the image-based
estimate of \Mbh{} and the one obtained from equation
\ref{eq_virial}, using FWHM instead of $v_{\rm BLR}$:
\begin{equation}\label{eq_f2}
f = \sqrt{\frac{{\cal M}_{\rm BH}({\rm image})}{{\cal M}_{\rm
BH}({\rm virial})}}
 = \sqrt{\frac{G {\cal M}_{\rm BH}({\rm image})}{R_{\rm BLR} {\rm FWHM}^2}}
\end{equation}

Tables \ref{tab_sample} and \ref{tab_spec} list the main measured and 
derived quantities for the blazar and reference samples. 

\section[]{Results and discussion}\label{sec_results}

\begin{figure}
\begin{center}
\includegraphics[width=0.49\textwidth]{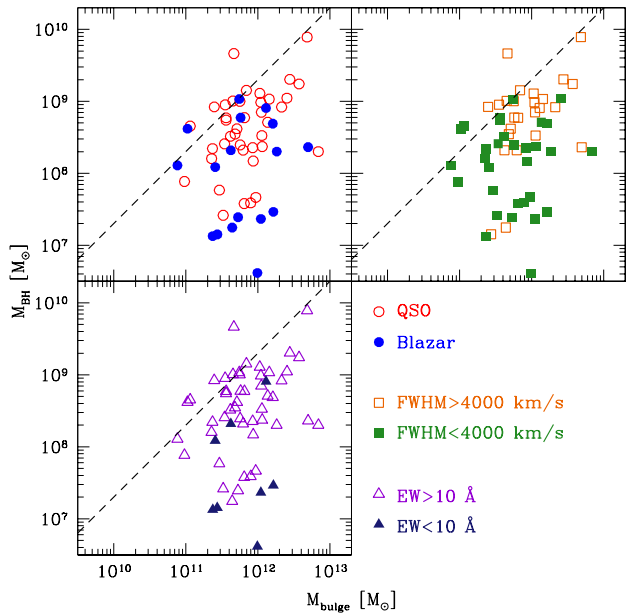}\\
\caption{Comparison between black hole masses (derived through the virial
technique, assuming FWHM=$v_{\rm BLR}$) and the mass of the host galaxy 
bulge. The local \citet{marconi03} relation is plotted as dashed lines. 
Different symbols mark different subsamples: Filled / empty circles mark 
the blazars / reference quasars. Filled / empty squares mark objects with 
line width smaller / larger than 4000 km/s. Filled / empty triangles mark 
objects with EW below / exceeding 10 \AA{}. Blazars (and BL Lac objects in 
particular) show smaller virial \Mbh{} than expected from the host galaxy 
mass, i.e., larger $f$ values.
}\label{fig_mbh}
\end{center}
\end{figure}

In Figure \ref{fig_mbh} we compare the \Mbh{} estimates obtained in the
virial assumption, assuming $f=1$, with the bulge mass inferred from the
host galaxy luminosity. It is apparent that blazars, and specifically 
BL Lac objects, show smaller virial \Mbh{} values than quasars for a 
given host galaxy mass, i.e., higher values of $f$ are required for blazars.
We note that the redshift-dependent evolution of the 
\Mbh{}/\Mhost{} scaling relation has not been applied at this stage.
The bulge mass range of blazars and quasars sampled in this study is the 
same, so no explicit bias due to the host galaxy is expected
to affect our results. The same is true if we divide our sample according 
to the broad line width: No obvious segregation in terms of bulge mass
is observed. Most importantly, when we pick up pure BL Lac objects
(with broad line Equivalent Width EW$<$10 \AA\footnote{Here we slightly
relaxed the usual EW$<$5 \AA{} threshold in order to include BL Lac and
3C279. We note that, in order to study broad emission lines, we are 
biased towards spectra collected during relatively faint states of the 
continuum emission of BL Lac objects.}), we automatically select blazars 
among those showing the largest deviations between \Mbh{} and 
${\cal M}_{\rm bulge}$, as expected if we assume that BL Lac objects show 
the smallest inclination angles amongst blazars.

\begin{figure}
\begin{center}
\includegraphics[width=0.49\textwidth]{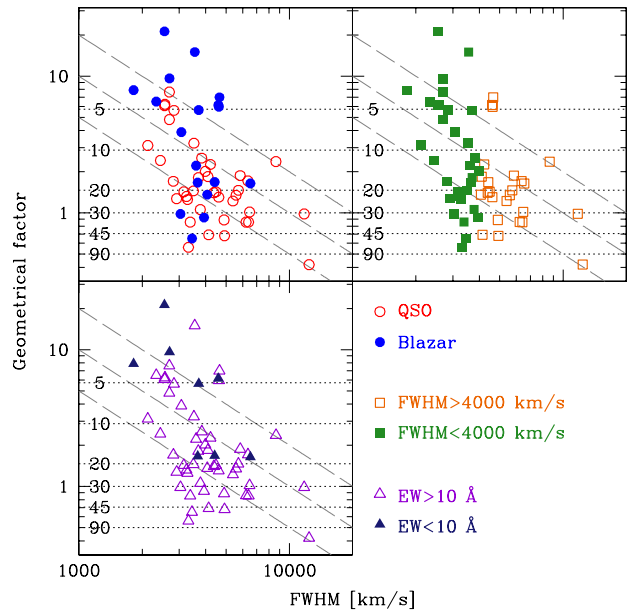}\\
\caption{The geometrical factor $f$ as a function of the
width of the lines (FWHM). Dotted lines show the expected values
of $f$ for a thin disc model of the BLR, with different inclination
angles (labelled in degrees). Diagonal, dashed lines mark the loci at
fixed cloud velocities in the thin disc model, namely $v_{\rm BLR}$=5000,
10000 and 20000 km/s. Symbol code is the same as in Figure \ref{fig_mbh}.
Blazars (and specifically BL Lac objects) show
relatively narrow lines and high $f$ values, consistently with the picture
of a flat BLR geometry observed pole-on.
}\label{fig_fwhmf}
\end{center}
\end{figure}

Figure \ref{fig_fwhmf} shows the $f$ values as a function of the line FWHM.
As already noted in \citet{decarli08}, an anti-correlation between
$f$ and FWHM is apparent. This supports the idea of a flat geometry viewed
at different inclination angles. In particular, only few objects with 
$f>2$ have FWHM$>$4000 km/s, and no one of them has FWHM$>$5000 km/s.
It is also apparent that blazars have, on average, higher $f$ than
quasars. Namely, $\langle f_{\rm BLLacs}\rangle=6.9\pm2.3$, 
$\langle f_{\rm blazars}\rangle=5.6\pm1.3$ and
$\langle f_{\rm quasars}\rangle=2.0\pm0.3$, with all the BL Lac objects
having $f\gsim2$. All these 
indications support a flat geometry of the BLR, with smaller inclination
angles for blazars and BL Lacs in particular.

\begin{figure}
\begin{center}
\includegraphics[width=0.49\textwidth]{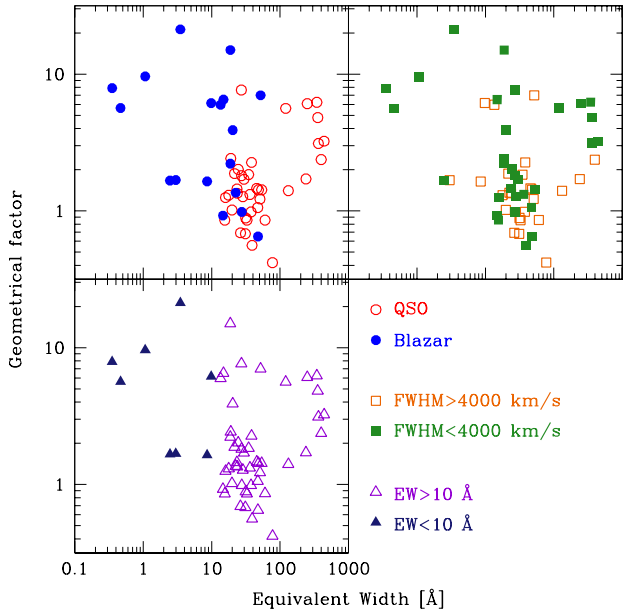}\\
\caption{The geometrical factor $f$ as a function of the Equivalent 
Width of broad lines. Symbol code is the same as in Figure
\ref{fig_mbh}. A trend is apparent, so that objects with strong jet
contamination (= small line Equivalent Width) all present large
geometrical factor values (consistent with a pole-on view of a flat
geometry).}\label{fig_ewf}
\end{center}
\end{figure}

Further evidence of this appears from the distribution of sources in
the $f$--EW plane. As shown in Figure \ref{fig_ll}, the luminosity ratio 
between broad lines and continua is almost constant in the average quasar 
population \citep[see also the analysis in][]{risaliti10}. In BL Lac
objects, the continuum emission from the jet contaminates the
optical spectrum, so that the ratio between the line flux and the
measured continuum (i.e., the line EW) is decreased. In Figure
\ref{fig_ewf} we show that a) no quasar is found with small EW, as
expected; b) the higher is $f$, the smaller is the EW. In
particular, all the objects with EW$<$10 \AA{} have $f\gsim2$. 
This picture is in good agreement with
the expectations from the unified model of AGN if we assume a flat 
geometry for the BLR.

\begin{figure}
\begin{center}
\includegraphics[width=0.49\textwidth]{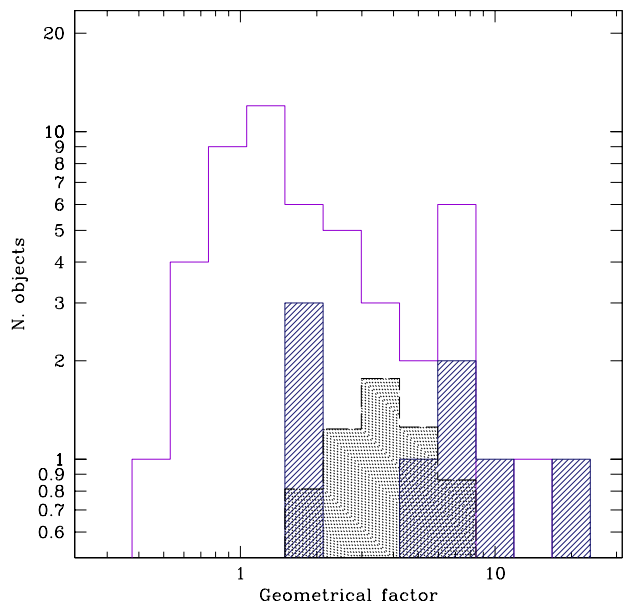}\\
\caption{Observed distributions of the geometrical factor
$f$ for objects with broad line Equivalent Widths exceeding / below 10 \AA{}
(solid empty / shaded histograms), as compared with the output
of our Montecarlo realizations for pole-on sources (dotted shaded 
histogram; see the text for details). The agreement is remarkable, 
given the modest statistics of BL Lac objects in our sample.
}\label{fig_fdistr}
\end{center}
\end{figure}
In order to put these results in the context of a quantitatively
defined picture of the BLR, we compare the distribution of $f$
observed in our study with the expectations from equation
\ref{eq_disc} through Montecarlo realizations. We generate a
compilation of $10^5$ objects with isotropically distributed
inclination angles $\vartheta$ between the BLR rotational axis and the
line of sight.  We consider as pole-on systems those having
$\vartheta<\vartheta_{\rm cut} = 10^\circ$. The cutoff angle is set by
the typical opening angle of a jet in BL Lac objects \citep[see][]{ghisellini93,urry95},
i.e., we select sources in which we expect the jet to contaminate the
continuum emission in the optical wavelengths.  Objects with
$\vartheta_{\rm cut}<\vartheta<\vartheta_{\rm max}\sim 40^\circ$ are
considered normal type-1 AGN. The cutoff between these two classes is
smoothed by adding an error $\sigma_\vartheta=10^\circ$ to the
intrinsic values of $\vartheta$. This uncertainty accounts for errors
in the estimate of $f$, intrinsic differences in the opening angles of
the jets, and physical mis--alignment between the jets and the BLR
rotational axis. This last effect is expected in a disc warped by,
e.g., the Bardeen--Petterson effect \citep[]{bardeen75}. In
particular, for very massive BHs as those discussed here, the warping
radius of a disc-like distribution of gas is comparable with the
radius of the BLR \citep[]{perego09}. In this picture, an average
misalignment of $\approx 10^{\circ}$ is expected if the BH is fuelled
by a larger scale circumnuclear disc \citep[]{dotti10}. Other physical
effects not considered here for sake of simplicity
\citep[e.g. self-irradiation driven warps, see][]{pringle96, maloney96}
could also play an additional role in smoothing the cutoff.

For each source we compute $f$ through equation \ref{eq_disc}, and we
compare the resulting distributions with what observed for objects with 
broad line Equivalent Width below 10 \AA{}. 
Expected and observed distributions of $f$ are shown in Figure 
\ref{fig_fdistr}, where we assumed $H/R=0$. Despite the modest number of 
BL Lac objects, the agreement between the predicted and observed distributions
is remarkably good. The Kolmogorov-Smirnov test indicates that the probability
that the observed distribution of $f$ for objects with EW$<$10 \AA{} is 
obtained by the same parent distribution that generated the pole-on sample
in the Montecarlo realization is high (40\%). 

Concerning $H/R$, the observation of few, very-high values of $f$ ($>10$)
suggests that, at least for these objects, $H/R\lsim0.05$. Re-running our
Montecarlo realizations with $H/R=0.05$ yields consistent results:
In particular, the probability that the pole-on sample and the observed BL 
Lac objects have the same parent distributions is $20\%$.

\section{Conclusions}\label{sec_conclusions}

We computed virial estimates of \Mbh{} in a set of 18 blazars and 39
reference quasars, and compared them with the \Mbh{} estimates derived
from the host galaxy luminosities. Virial black hole masses are derived
through broad line luminosities and widths, using a newly defined relation
between the line and the continuum luminosity in quasars. From the 
comparison of the two \Mbh{} estimates we infer the geometrical factor
$f$, responsible for the deprojection of the virial velocity from the
line of sight, and we used $f$ to constrain the properties of the broad
line region in blazars. We found that:
\begin{itemize}
\item[{\it i-}] Blazars, and BL Lac objects in particular, have higher
$f$ values on average with respect to the reference quasar sample, and
both these classes have average $f$ values well exceeding the predictions
for an isotropic BLR. The average values of $f$ for each class of objects
are: $\langle f_{\rm BLLacs}\rangle=6.9\pm2.3$, 
$\langle f_{\rm blazars}\rangle=5.6\pm1.3$ and
$\langle f_{\rm quasars}\rangle=2.0\pm0.3$.
\item[{\it ii-}] The geometrical factor clearly depends on the width of
broad lines, in the sense that only few objects with $f\gsim 2$ 
have FWHM$>$4000 km/s, and no one of them has FWHM$>$5000 km/s. All the 
blazars in our sample have relatively narrow lines ($\lsim$5000 km/s).
\item[{\it iii-}] An anti-correlation between the geometrical factor 
and the broad line equivalent width is also found, in the sense that
all the objects with EW$<$10 \AA{} have geometrical factors $f\gsim 2$.
\end{itemize}
These results rule out an isotropic sketch of the broad line region.
The observed distribution of $f$ is in excellent agreement with
the expectations for a geometrically-thin ($H/R\lsim 0.05$) disc picture.
In this scheme, the broad line region of BL Lac objects is seen almost
pole-on. Assuming small or negligible tilts ($<$ few degrees) between the 
disc rotational axis and the AGN jet, this conclusion is in full agreement 
with the Unified Model of AGN.

\section*{Acknowledgments}
We are grateful to the Referee for his/her prompt feedback and fruitful 
comments. We thank G. Ghisellini for useful discussions.

%
%

\bsp

\label{lastpage}


\begin{thebibliography}{99}
 \bibitem[\protect\citeauthoryear{Abdo et al.}{2009a}]{abdo09a} Abdo A.A., Ackermann M., Ajello M., Axelsson M., Baldini L., Ballet J., Barbiellini G., Bastieri D., et al., 2009a, ApJ, 707, 727
 \bibitem[\protect\citeauthoryear{Abdo et al.}{2009b}]{abdo09b} Abdo A.A., Ackermann M., Ajello M., Baldini L., Ballet J., Barbiellini G., Bastieri D., Bechtol K., et al., 2009b, ApJ Letters, 707, 142
 \bibitem[\protect\citeauthoryear{Antonucci \& Miller}{1985}]{antonucci85} Antonucci R.R.J. \& Miller J.S., 1985, ApJ, 297, 621
 \bibitem[\protect\citeauthoryear{Bardeen \& Petterson}{1975}]{bardeen75} Bardeen J. M., Petterson J. A., 1975, ApJ, 195, L65
 \bibitem[\protect\citeauthoryear{Bentz et al.}{2009}]{bentz09} Bentz M.C., Peterson B.M., Netzer H., Pogge R.W., Vestergaard M., 2009, ApJ, 697, 160
 \bibitem[\protect\citeauthoryear{Bettoni et al.}{2003}]{bettoni03} Bettoni D., Falomo R., Fasano G. \& Govoni F., 2003, A\&A, 399, 869
 \bibitem[\protect\citeauthoryear{Brotherton}{1996}]{brotherton96} Brotherton M.S., 1996, ApJS, 102, 1
 \bibitem[\protect\citeauthoryear{Bruzual \& Charlot}{2003}]{bruzual03} Bruzual G., Charlot S., 2003, MNRAS, 344, 1000
 \bibitem[\protect\citeauthoryear{Capetti, Raiteri \& Buttiglione}{2010}]{capetti10} Capetti A., Raiteri C.M., Buttiglione S., 2010, A\&A, 516, 59
 \bibitem[\protect\citeauthoryear{Collin et al.}{2006}]{collin06} Collin S., Kawaguchi T., Peterson B.M., Vestergaard M., 2006, A\&A, 456, 75
 \bibitem[\protect\citeauthoryear{Decarli et al.}{2008a}]{nls1} Decarli R., Dotti M., Fontana M., Haardt F., 2008a, MNRAS Letters, 386, 15
 \bibitem[\protect\citeauthoryear{Decarli et al.}{2008b}]{decarli08} Decarli R., Labita M., Treves A., Falomo R., 2008b, MNRAS, 387, 1237
 \bibitem[\protect\citeauthoryear{Decarli et al.}{2010a}]{decarli10a} Decarli R., Falomo R., Treves A., Kotilainen J.K., Labita M., Scarpa R., 2010a, MNRAS, 402, 2441
 \bibitem[\protect\citeauthoryear{Decarli et al.}{2010b}]{decarli10b} Decarli R., Falomo R., Treves A., Labita M., Kotilainen J.K., Scarpa R., 2010b, MNRAS, 402, 2453
 \bibitem[\protect\citeauthoryear{De Rosa et al.}{2010}]{derosa10} De Rosa G., Decarli R., Walter F., Fan X., Jiang L., Kurk J., Pasquali A., Rix H.W., 2010, ApJ submitted
 \bibitem[\protect\citeauthoryear{Dotti et al.}{2010}]{dotti10} Dotti M., Volonteri M., Perego A., Colpi M., Ruszkowski M., Haardt F., 2010, MNRAS, 402, 682
 \bibitem[\protect\citeauthoryear{Fine et al.}{2008}]{fine08} Fine S., Croom S.M., Hopkins P.F., Hernquist L., Bland-Hawthorn J., Colless M., Hall P.B., Miller L., et al., 2008, MNRAS, 390, 1413
 \bibitem[\protect\citeauthoryear{Fine et al.}{2010}]{fine10} Fine S., Croom S.M., Bland-Hawthorn J., Pimbblet K.A., Ross N.P., Schneider D.P., Shanks T., 2010, MNRAS, accepted (arXiv:1005.5287)
 \bibitem[\protect\citeauthoryear{Ghisellini et al.}{1993}]{ghisellini93} Ghisellini G., Padovani P., Celotti A., Maraschi L., 1993, ApJ, 407, 65
 \bibitem[\protect\citeauthoryear{Grupe \& Mathur}{2004}]{grupe04} Grupe D., Mathur S., 2004, ApJ Letters, 606, 41
 \bibitem[\protect\citeauthoryear{Kaspi et al.}{2000}]{kaspi00} Kaspi S., Smith P.S., Netzer H., Maoz D., Jannuzi B.T., Giveon U., 2000, ApJ, 533, 631
 \bibitem[\protect\citeauthoryear{Kaspi et al.}{2005}]{kaspi05} Kaspi S., Maoz D., Netzer H., Peterson B.M., Vestergaard M., Jannuzi B.T., 2005, ApJ, 629, 61
 \bibitem[\protect\citeauthoryear{Kaspi et al.}{2007}]{kaspi07} Kaspi S., Brandt W.N., Maoz D., Netzer H., Schneider D.P., Shemmer O., 2007, ApJ, 659, 997
 \bibitem[\protect\citeauthoryear{Kotilainen et al.}{1998}]{kotilainen98} Kotilainen J.K., Falomo R., \& Scarpa R., 1998, A\&A, 332, 503
 \bibitem[\protect\citeauthoryear{Labita et al.}{2006}]{labita06} Labita M., Treves A., Falomo R., Uslenghi M., 2006, MNRAS, 373, 551
 \bibitem[\protect\citeauthoryear{La Mura et al.}{2007}]{lamura07} La Mura G., Popovi\'c L.\v{C}, Ciroi S., Rafanelli P., Ili\'c D., 2007, ApJ, 671, 104
 \bibitem[\protect\citeauthoryear{La Mura et al.}{2009}]{lamura09} La Mura G., Di Mille F., Ciroi S., Popovi\'c L.\v{C}, Rafanelli P., 2009, ApJ, 693, 1437
 \bibitem[\protect\citeauthoryear{Maloney, Begelman \& Pringle}{1996}]{maloney96} Maloney P.R., Begelman M.C., Pringle J. E., 1996, ApJ, 472, 582
 \bibitem[\protect\citeauthoryear{Mannucci et al.}{2001}]{mannucci01} Mannucci F., Basile F., Poggianti B. M. et al., 2001, MNRAS, 326, 745
 \bibitem[\protect\citeauthoryear{Marconi \& Hunt}{2003}]{marconi03} Marconi A., Hunt L.K., 2003, ApJ Letters, 589, 21
 \bibitem[\protect\citeauthoryear{McLure \& Dunlop}{2001}]{mclure01} McLure R.J.,  Dunlop J.S., 2001, MNRAS, 327, 199
 \bibitem[\protect\citeauthoryear{McLure \& Jarvis}{2002}]{mclure02} McLure R.J. \& Jarvis M.J., 2002, MNRAS, 337, 109
 \bibitem[\protect\citeauthoryear{Netzer}{1990}]{netzer90} Netzer H., 1990, in Active Galactic Nuclei, ed. R. D. Blandford, H. Netzer, \& L. Woltjer (Berlin: Springer), 137
 \bibitem[\protect\citeauthoryear{Nilsson et al.}{2003}]{nilsson03} Nilsson K., Pursimo T., Heidt J., Takalo L.O., Sillanp\"a\" A., Brinkmann W., 2003, A\&A, 400, 95
 \bibitem[\protect\citeauthoryear{Nilsson et al.}{2009}]{nilsson09} Nilsson K., Pursimo T., Villforth C., Lindfors E., Takalo L.O., 2009, A\&A, 505, 601
 \bibitem[\protect\citeauthoryear{Nilsson et al.}{2010}]{nilsson10} Nilsson K., Takalo L.O., Lehto H.J., Sillanp\"a\"a A., 2010, A\&A, 516, 60
 \bibitem[\protect\citeauthoryear{Onken et al.}{2004}]{onken04} Onken C.A., Ferrarese L., Merritt D., Peterson B.M., Pogge R.W., Vestergaard M., Wandel A., 2004, ApJ, 615, 645
 \bibitem[\protect\citeauthoryear{Perego et al.}{2009}]{perego09} Perego A., Dotti M., Colpi M., Volonteri M., 2009, MNRASm 399, 2249
 \bibitem[\protect\citeauthoryear{Pian, Falomo \& Treves}{2005}]{pian05} Pian E., Falomo R., Treves A., 2005, MNRAS, 361, 919
 \bibitem[\protect\citeauthoryear{Pringle}{1996}]{pringle96} Pringle J.E., 1996, MNRAS, 281, 357
 \bibitem[\protect\citeauthoryear{Risaliti, Salvati \& Marconi}{2010}]{risaliti10} Risaliti G., Salvati M., Marconi A., 2010, MNRAS accepted (arxiv:1010.2037)
 \bibitem[\protect\citeauthoryear{Sbarufatti et al.}{2005}]{sbarufatti05} Sbarufatti B., Treves A., Falomo R., Heidt J., Kotilainen J., Scarpa R., 2005, AJ, 129, 559
 \bibitem[\protect\citeauthoryear{Sbarufatti et al.}{2006a}]{sbarufatti06aa} Sbarufatti B., Falomo R., Treves A., Kotilainen J., 2006, A\&A, 457, 35
 \bibitem[\protect\citeauthoryear{Sbarufatti et al.}{2006b}]{sbarufatti06aj} Sbarufatti B., Treves A., Falomo R., Heidt J., Kotilainen J., Scarpa R., 2006b, AJ, 132, 1
 \bibitem[\protect\citeauthoryear{Scarpa et al.}{2000}]{scarpa00} Scarpa R., et al., 2000, ApJ, 532, 740
 \bibitem[\protect\citeauthoryear{Shen et al.}{2010}]{shen10} Shen Y., Hall P.B., Richards G.T., Schneider D.P., Strauss M.A., Snedden S., Bizyaev D., Brewington H., et al. 2010, arXiv:1006.5178
 \bibitem[\protect\citeauthoryear{Urry \& Padovani}{1995}]{urry95} Urry C.M. \& Padovani P., 1995, PASP, 107, 803
 \bibitem[\protect\citeauthoryear{Vestergaard, Wilkes \& Barthel}{2000}]{vestergaard00} Vestergaard M., Wilkes B.J., Barthel P.D., 2000, ApJ, 538L, 103
 \bibitem[\protect\citeauthoryear{Vestergaard \& Wilkes}{2001}]{vestergaard01} Vestergaard M., Wilkes B.J., 2001, ApJ Suppl., 134, 1
 \bibitem[\protect\citeauthoryear{Vestergaard \& Peterson}{2006}]{vestergaard06} Vestergaard M., \& Peterson B.M., 2006, ApJ, 641, 689
 \bibitem[\protect\citeauthoryear{Wills \& Browne}{1986}]{wills86} Wills B.J. \& Browne I.W.A., 1986, ApJ, 302, 56
 \bibitem[\protect\citeauthoryear{Wurtz et al.}{1996}]{wurtz96} Wurtz, Stocke \& Yee, 1996, ApJS, 103, 109
 \bibitem[\protect\citeauthoryear{Yuan et al.}{2008}]{yuan08} Yuan W., Zhou H.Y., Komossa S., Dong X.B., Wang T.G., Lu H.L., Bai J.M., 2008, ApJ, 685, 801
\end{thebibliography}
\end{document}